# Highly Efficient Second Harmonic Generation of Thin Film Lithium Niobate Nanograting near Bound States in the Continuum

Zhijin Huang [1], Mengjia Wang [2], Yang Li [1], Jumei Shang [1], Ke Li [1], Wentao Qiu [1], Jiangli Dong[3], Heyuan Guan [3*], Zhe Chen [1], and Huihui Lu[1*]

[1] *Key Laboratory of Optoelectronic Information and Sensing Technologies of Guangdong Higher Education Institute, Jinan University, Guangzhou 510532, P.R. China*

[2] *FEMTO-ST Institute UMR 6174, University of Bourgogne Franche-Comte CNRS, Besancon 25030, France*

[3] *Guangdong Provincial Laboratory of Optical Fiber Sensing and Communications, Jinan University, Guangzhou 510632, China;*

*Corresponding author: H. Guan: ttguanheyuan@jnu.edu.cn; H. Lu: thuihuilu@jnu.edu.cn

Bound states in the continuum (BICs), a concept from quantum mechanics, are ubiquitous physical phenomena where waves will be completely locked inside physical systems without energy leaky. Such a physical phenomenon in optics will provide a platform for optical mode confinement to strengthen local field enhancement in nonlinear optics. Here we utilize an optical system consisting of asymmetric nanograting and waveguide of thin film lithium niobate (LiNbO$_3$) material to enhance second harmonic response near BICs. By breaking the symmetry of grating periodicity, we realize strong local field confined inside waveguide up to 35 times normalized to incident field (with dissymmetric factor δ of 0.1), allowing strong light-matter interaction in nonlinear material. From the numerical simulation, we theoretically demonstrate that such an optical system can greatly enhance second harmonic intensity enhancement of ~10$^4$ compared with undersigned LiNbO$_3$ film and conversion efficiency reaching 1.53×10$^{-5}$ at δ =0.2 under illumination of 1.33 GW/cm$^2$. Surprisingly, we can predict that a giant enhancement of second harmonic conversion efficiency will exceed 8.13×10$^{-5}$ at δ =0.1 when the optical system is extremely close to BICs. We believe that such an optical system to trap local field inside is also accessible to promote the application of thin film lithium niobate in the field of integrated nonlinear optics.

**Lithium Niobate, Second harmonic generation, Bound States in the continuum**

## 1 Introduction

Bound states in the continuum (BICs) are ubiquitous physical phenomena, which were originally proposed in hypothetical quantum system by Neumann and Wigner [1–3]. It is a general wave phenomenon that has been not only found in many physical systems including photonic crystal [4],optical grating and waveguide[5,6], metasurfaces[7,8], but also have been utilized to realized some interesting physical phenomena such as diffraction-free beam guiding beyond light cone[9] and the enhancement of Goos-Hänchen Shift[10]. Spectrally, BICs are the points where two resonant waves interfere, leaving one resonance with the quality factor (Q-factor) diverging to infinity. By themselves, the BICs are localized solutions decoupled from any external wave's incident on the system. However, even for the slightest off-set from the BICs in the momentum space transformations, the high-Q resonant modes with unlimited Q-factors with lossless materials can be configured in the optical systems [11,12]. In this aspect, a related point is the maximal achieved field confinement near the BICs, which is called quasi-embedded bond-state resonance configuration occurring between highly leaky and completely trapped mode conditions [4,13]. To observe the appearance of ultrahigh-Q resonances in lossless optical systems, one way is to break the symmetry of optical system to realize extremely narrow resonance, also called trapped modes [4,12]. The excitation of the strong trapped mode results in strong field enhancement in optical systems is a dominant feature of light-matter interaction.

Second harmonic generation(SHG) is an essential technology, doubling the frequency of light by interaction with nonlinear material[15], which enjoys widespread applications including holographic imaging[16], human face and QR code recognitions[17] and harmonic optical tomography[18], etc. However, since the nonlinear optical response is intrinsically weak, advances in artificial nanostructured plasmonic and all-dielectric metasurfaces have enabled nonlinear optical processes that do not satisfy phase matching condition strictly[19–21]. These techniques are created to realize tight confinement and large local field enhancement of electromagnetic fields in a small volume, which generate much higher nonlinear conversion efficiencies than in the constituent materials[22]. Moreover, these techniques pave the way for a wave of revolutionary concepts and devices that will work more efficient and flexible harmonic generation or frequency mixing. Theoretically, the second-order nonlinear polarization can be described by $P^{(n)}(\omega) \propto \chi^{(n)}(\omega,\omega)[E_{loc}(\omega)]^n$. Following this equation, it is very clear that the nonlinear signals are determined by the local-field intensity at fundamental frequency. As a result, enormous efforts have been poured into boosting the field enhancement. However, developments of resonant dielectric structures with high-Q resonance driven by BICs has been observed in boosting second harmonic generation[23,24], for example, for example, second- and third-harmonic generation from periodic arrangement of subwavelength resonators with high-refractive index materials such as Si, GaAs, supporting sharp high-Q resonance based on symmetry-protected BICs. However, among nonlinear materials, lithium niobate (LN) is an excellent choice for nonlinearly light generation due to its large second-order susceptibility ($d_{33}$~27pm/V)[25] and wide applications in field of electro-optical modulators. And it is a low-loss material with a transparency from visible to infrared regime. As a result, enormous efforts have been poured into boosting the SH conversion efficiency, for example optical anapole mode assisted SHG in LiNbO$_3$ nanodisk[26,27], gold nanoring embedded LiNbO$_3$ nanocylinder for SHG enhancement[28], Fano resonant couple systems for high efficient SHG[29]. In this sense, BICs contributing to strong light-matter interaction provides a platform for generating second harmonic waves.

Here, we demonstrate a high Q-factor resonant optical system consisting of a nanograting and waveguide of x-cut LiNbO$_3$ placed on a fused silica substrate. And we begin by presenting the design principle and numerical simulation to reveal the physical mechanic of BICs in the LNGW optical system. Next, we exploit the reflection spectra and local field confinement embedded in LNGW by shifting the dissymmetry factor δ from -1 to 1. Ultrahigh Q-factor resonance at quasi-BICs achieves giant local field enhancement, which open an avenue for boosting light-matter interaction in second harmonic generation. Furthermore, we utilize this mechanic to improving the second-order response of LNGW structure and demonstrate the dependence between SH conversion efficiency and the dissymmetry factor of the structure detailed. Finally, we conclude our findings and show that the optical system can be utilized in nonlinear optical wave generation. In this paper, the ordinary and extraordinary refractive index of LiNbO$_3$ is referred to Ref[25].

## 2 BICs in LiNbO$_3$ Waveguide-Grating Optical System

In this section, theoretical and numerical calculation is utilized to go deep into the striking physical phenomenon of BICs. We start with an optical system consisting of 360 nm-thick LiNbO$_3$ waveguide and an asymmetric LiNbO$_3$ grating with 290 nm in thickness and 400 nm in grating periodicity placed on SiO$_2$ substrate in free space, illuminated oblique with z-polarized (TE-polarized) plane wave as illustrated in Fig.1 (a). The extraordinary refractive index ($n_e$~2.2 at 690 nm) will be taken into consideration in this paper. For the sake of simplicity, we utilize Helmholtz equation to calculate the dispersion diagrams of the TE-polarized mode inside LiNbO$_3$ waveguide as follow:

$$\frac{d^2E_z}{dx^2} + \left[k_0^2 n^2 - \beta^2\right]E_z = 0 \quad (1)$$

Here $n$ is the refractive index of LN; $k_0$ is the incident wavevector; $\beta$ is propagation constant. By considering the boundary condition of LN waveguide, we can further obtain the Eigen mode by solving Eq.1:

$$\frac{\omega h}{2\pi c} = \frac{1}{n_{wg}}\sqrt{\left(\frac{\beta h}{2\pi}\right)^2 + \left(\frac{\varphi_0 + m\pi}{2\pi}\right)^2} \quad (2)$$

Here $\varphi_0$ is phase constant; $m$=0, 1, 2, corresponds to TE$_0$, TE$_1$, TE$_2$ mode, respectively. Only when the waveguide's propagation constant satisfies: $k_0 n_1 > \beta > k_0 n_2$, we obtain discrete guided modes shown in Fig.1 (b). As for the thin film LN nanograting, the diffraction wave vector is determined by:

$$\frac{k_{x,n}h}{2\pi} = \frac{k_0 h \sin(\theta)}{2\pi} - n\frac{h}{\Lambda} \quad (n = 0, \pm 1, \pm 2, \ldots) \quad (3)$$

The incident angle is set to be $\theta$=5°, $n$ is the order of diffraction wave vector, $k_0$ is wave vector in free space.
To illustrate further this physical phenomenon, we plot the dispersion relationship of waveguide modes (TE$_0$, TE$_1$, TE$_2$, representing by blue, orange and yellow line, respectively) and grating evanescent order m=-1 (pink dashed line) to-gether, as shown in Fig.1 (b). The longitudinal axis represents normalized frequency while the transverse axis is normalized wavevector (k=2pi/λh). It is clear that two intersection points between the evanescent order m=-1 of LiNbO$_3$ grating and TE$_1$ mode of LiNbO3 waveguide, which are marked as red circle at 3.5. And it indicates that BICs will occur at these two points due to the interference between waveguide mode and grating mode. To further investigate the BICs by tuning parameters of LNGW structure, reflection spectra were carried out by FDTD simulation with perfect matched layer (PML) in x-direction and periodic boundary condition (PBC) in y-axis by commercial FDTD software. The oblique incident light is TE-polarized with electric field parallel z-axis in Cartesian coordinate. For the sake of simplicity, we assume that the LNGW optical system consists of two grating with periodicities Λ$_1$ and Λ$_2$. So, we can define the dissymmetric factor as follows:

$$k'_{x,m} = nk_0 \sin\theta - m\pi\left(\frac{1}{\Lambda_1} + \frac{1}{\Lambda_2}\right) \quad (4)$$

$$\beta = k_{x,m} = nk_0 \sin\theta - m\pi\left(\frac{1}{\Lambda}\right) \quad (5)$$

$$\delta = \left(\frac{2\Delta d}{\Lambda - 2w}\right) \quad (6)$$

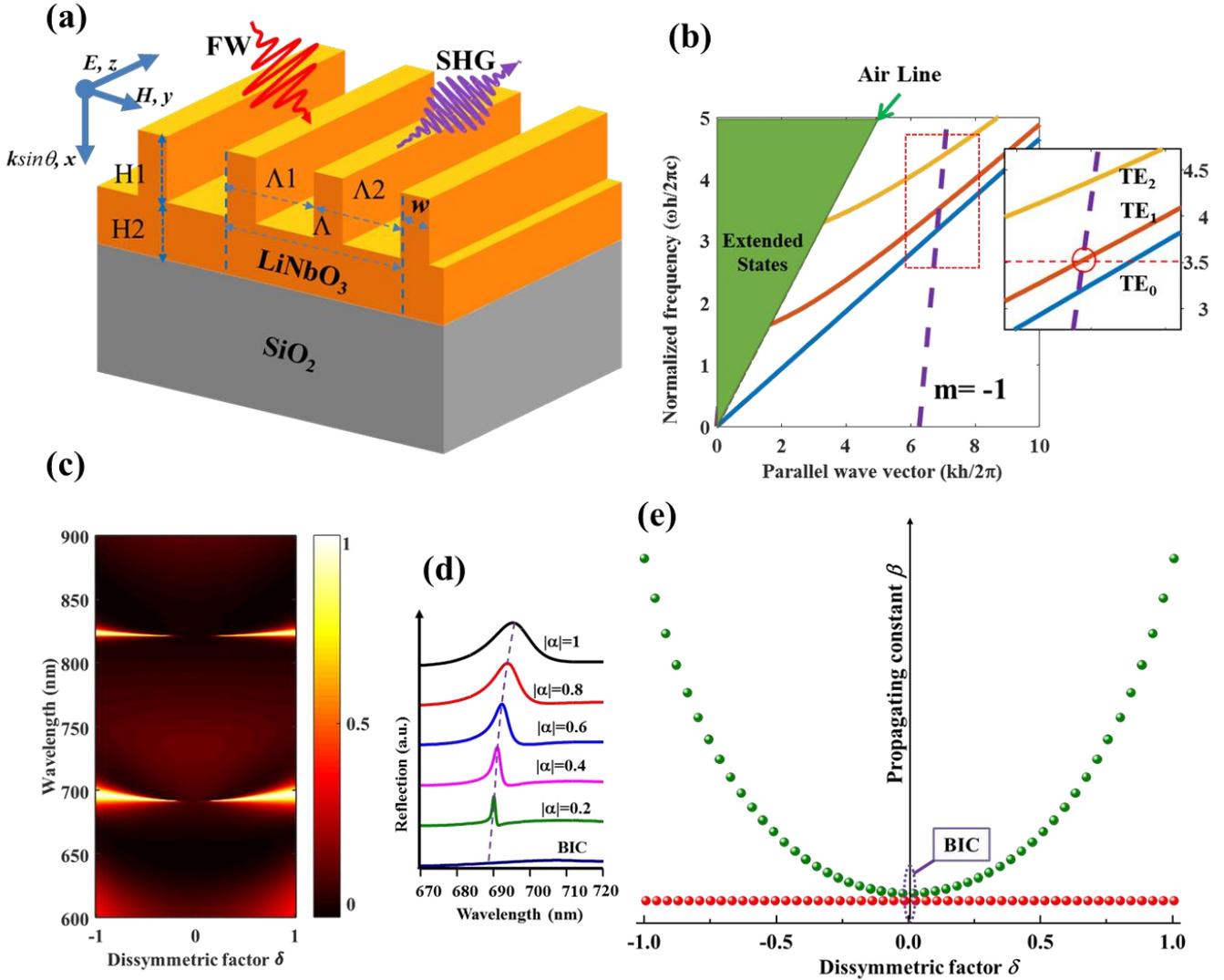

**Figure.1**. Optical characteristics of the LNWG optical systems. (a) Schematic illustration of LN waveguide-grating optical system, consisting of LiNbO$_3$ waveguide (360nm in thickness) and asymmetric gratings (290nm in thickness). The dissymmetric factor is defined as δ= 2Δd/ (Λ-2w) varying from -1 to 1. (b) Harmonic mode frequency for LN waveguide (n$_e$=2.2). Orange lines correspond to modes that are localized in the LN waveguide. The shaded blue region is a continuum states that extend into both SiO2 substrate and air around it. The pink lines represent the diffraction dispersion states of LN nanograting. Here the first evanescent order is taken into consideration. (c) Numerical calculation of reflection spectra with dissymmetric factor diverse from -1 to 1. It sees that the spectra will disrupt at wavelengths of approximately ∼690nm and ∼820nm when δ is zero. This case is called BICs. (d) Transmission spectra with various δ in visible regime from 670 nm to 720 nm. (e) Mode-coupling analysis between gratings and waveguide, marked by red spheres and olive spheres, respectively. BICs occurs when two modes satisfy phase matching condition

Where $n$ is the refractive index of air, $k_0$ is the wavevector of free-space light and $\theta$ is the angle of illuminating on the LiNbO$_3$ nanograting, $\Delta d$ is shift distance of nanograting and $2\Delta d = \Lambda_1 - \Lambda_2$. Equation (4) and (5) represent the arbitrary wavevector generated by the grating and wavevector satisfying mode-matched condition, respectively. To considering symmetry-protected bound states in the continuum, we define the dissymmetric factor by equation (6), varying from -1 to 1. As illustrated in Fig.1(c), two breaking points occur at a wavelength of 600 nm to 820 nm when the structure is symmetric. As the grating periodicity $\Lambda_1$ or $\Lambda_2$ is slightly changed, ultra-narrow asymmetric Fano-line-shape resonance occurs due to asymmetric coupling and interference between grating diffractive mode and waveguide discrete mode, which are similar to the continuum state and discrete state in quantum mechanics. To further investigate the coupling process, we calculate the waveguide constant and grating diffractive wavevector with Eq. (4) and (5) together as a function of dissymmetric factor $\delta$. As the results is shown in fig.1 (e), the BICs will occur when the grating diffractive wave vector $k_{x,m}$ (olive spheres) match with the waveguide propagating constant $\beta$ (red spheres), meaning the phase difference is decreased to zero where the optical mode is locked in inside the waveguide without any radiation. When the dissymmetric factor is nonzero, the phase difference is no longer zero, leading to asymmetric ultra-sharp line shape

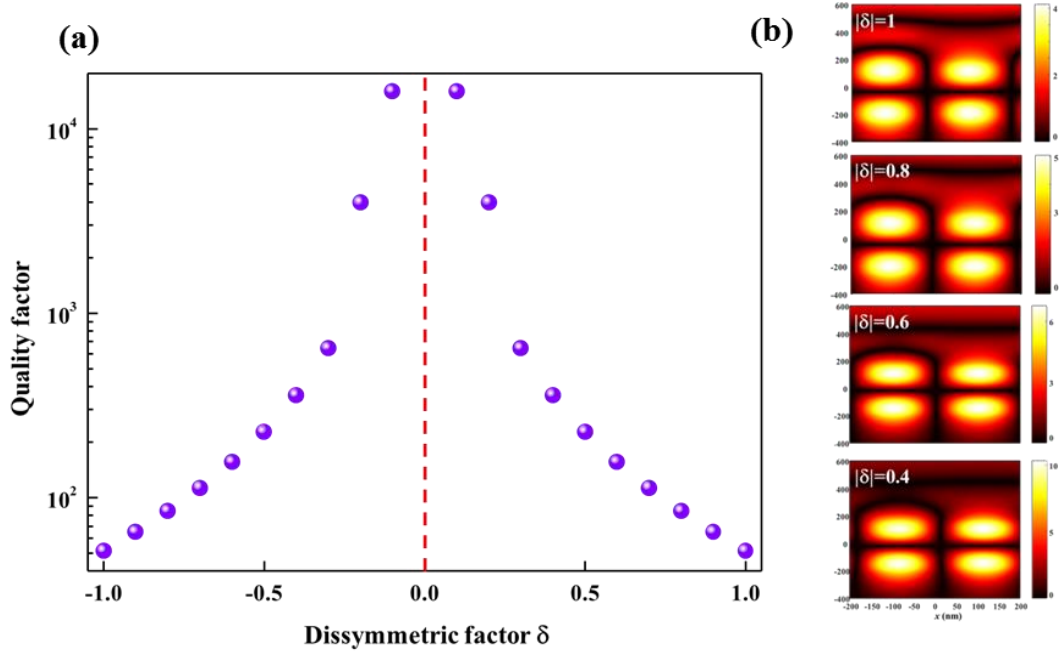

**Figure.2**. (a) Log-plot Quality factor along with the dissymmetric factor. The Q factor has a magnitude of $10^4$ at $|\delta|=0.1$ and diverges to infinity at BICs ($\delta=0$, marked by red dashed line). (b) Local field distribution increases ($|E/E_0|$) at $|\delta|=1, 0.8, 0.6$ and $0.4$. More clearly, the optical field is well confined inside the LiNbO$_3$ waveguide rather than the asymmetric gratings.

The quality factor (Q factor) is defined as[30]

$$Q = \left(\frac{f_R}{FWHM}\right) \quad (7)$$

Where $f_R$ is the resonant frequency and FWHM is the full width half max of the resonance intensity spectrum. The Q factor goes up to infinity as $\delta$ extremely approaches to zero, i.e., BICs. This case means optical mode is completely confined inside LNWG structure without any energy leaky.

Fig.2 (b) shows the local field distribution at $|\delta|=1, 0.8, 0.6, 0.4$, indicating that strong local field confinement will be realized when parameters of optical system is tuned extremely close to BICs. In this case, the Q factor is still very large, up to $10^4$ but optical energy will leak from the waveguide, that is so called the quasi bound states in the continuum, and it is very significant to enhance light-matter interaction in nonlinear optics or the observation of embedded striking physical phenomena.

## 3 Giant SH Response Assisted quasi-BIC

According to the previous work[31–33], strong local optical field confinement plays a crucial important role in giant nonlinear response in the field of micro and nano structures. Here, we first investigate the second harmonic generation (SHG)

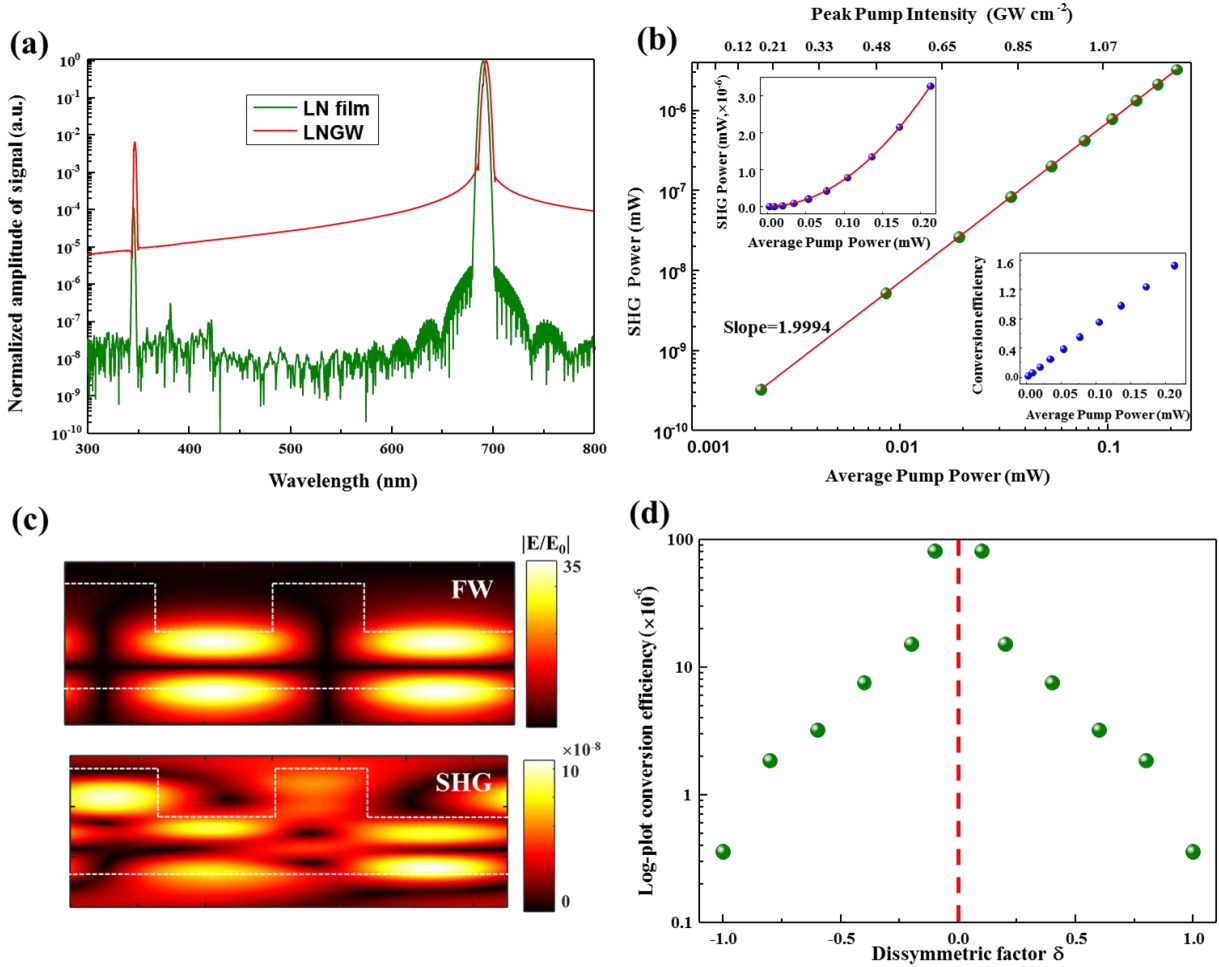

Figure3. Characteristics of second harmonic wave generated by LNGW optical system at dissymmetric factor $|\delta|=0.1$. (a) Normalized amplitude of SHG signal generated from LNGW (red line) and LN thin film (olive line), respectively. All the excitation intensity is ~1.33 GW/cm² with a wavelength of ~690 nm. (b) Log-plot power dependence for the SHG signals of LNGW structure under the illumination from 0.002 mW to 0.22 mW with a fitting slope f 1.9994. The inset at right corner shows that SHG conversion efficiency increases as well as average pump power increases. When the peak intensity of pump light is 1.33 GW/cm², corresponding to average power of 0.22 mW, the SHG conversion efficiency reaches to $1.53\times10^{-5}$. (c) Normalized optical field distribution at fundamental wavelength of 690 nm and second harmonic wavelength of 345 nm with $|\delta|=0.1$. (d) SHG conversion efficiency shows inseparable dependence with the dissymmetric factor δ. Clearly, SHG conversion efficiency ($|\delta|=0.1$, marked by red dashed circle) will greatly boost extremely close to BICs marked red dashed line.

of LNGW structure at dissymmetric factor δ=0.2 with numerical simulation by means of FDTD solution for the sake of simplicity. And the case of δ=0.1 is also taken into consideration to make a comparison. The SH signal is calculated by integrating the energy flux passing through the field-power monitor[34]. Based on the literatures[35–38], the maximum second-order susceptibility of lithium niobate is $d_{33}$~27pm/V. In the SHG simulations, the fundamental excitation power of the light source, the pulse duration and the repetition rate are selected as 0.001 mW to 0.22 mW, 200 fs, and 80 MHz, respectively. Figure3 shows the calculated SHG characteristics of LNGW structure. In fig.3 (a), the red and olive lines represent the calculated SHG signals of LNWG structure and LN thin film respectively under fundamental excitation wavelength of ~690 nm (SHG signal~345 nm) with the averaged intensity of 1.33 GW/cm², corresponding to an amplitude of $1\times10^8$ V/m. Compared to undersigned thin film LN structure, our proposed structure demonstrate a SHG amplitude of ~100 times larger, equivalent to an intensity enhancement of ~$10^4$ in magnitude. For practical application, it is very interesting to investigate the averaged power dependence of the SHG signal for our proposed structure; we plot the SHG power as a function of the average power of excited light from 0.002 mW to 0.22 mW corresponding a peak intensity varying from 0.12 to 1.33 GW/cm², as shown in Fig.3 (b). Notably, the SHG power possesses a squared relationship with the increasing excited average pump power (in the left inset), showing a linear trend with a slope of ~2 in the log plot. The results agree well with the equation[15,39]

$$\log(P_{avg}^{SH}) \propto 2\log(P_{avg}^{FW}) \ (8)$$

Here, $P_{avg}^{SH}$ is the second harmonic output power, and $P_{avg}^{FW}$ is the power of excited source. By dividing total SHG power by the FW power to define the conversion efficiency

$$\eta = P_{avg}^{SH} / P_{avg}^{FW} \ (9)$$

We calculate the value up to $1.53\times10^{-5}$ with the averaged excited power of 0.22 mW and a peak intensity of 1.33GW/cm² when the dissymmetric actor|δ|=0.2 in the right inset of Fig.3 (b). However, we follow the figure of merit ζ Eq.10 to theoretically quantify the SHG process in this paper as follows:

$$\xi = P_{peak}^{SH} / \left(P_{peak}^{FW}\right)^2 \ (10)$$

Here, $P_{peak}^{SH}$ and $P_{peak}^{FW}$ are the peak power at second harmonic and fundamental wavelength. This FOM is independent of the power of pump source, which provides a consistent and comparable parameter with other structure[40]. Thus, we calculate the FOM up to $9.18\times10^{-6}$ W⁻¹ at δ=0.2 with pump peak intensity of 1.33 GW/cm⁻². From the linear and nonlinear local field distribution at δ=0.2, this result also proves that the excited intensity has not reached saturation yet when the dissymmetric factor is ~0.2 (i.e., far from quasi BICs), the giant SHG signal of LNGW structure can be effectively excited when closely to BICs because optical energy locked inside nonlinear material will occur at these points, as shown in Fig.3 (d). To confirm this view, we calculate the SH conversion efficiency is up to $8.13\times10^{-5}$ in the case where dissymmetric factor |δ|=0.1, which is 5 times higher that of the case |δ|=0.2 while the FOM reaches $1.08\times10^{-5}$ W⁻¹. In table 1, it indicates that our proposed structure has a good performance in improving SH conversion efficiency. At this point, nonlinear wave conversion efficiency will be boosted up to a higher level when the dissymmetric factor δ is extremely close to zero.

*Table1. Comparison of SHG conversion efficiency at bound state in the continuum*

| Structure type | Pump power/ peak intensity | $\eta = P^{SH}/P^{FW}$ | $\eta = P^{SH}/\left(P^{FW}\right)^2$ (W⁻¹) |
|---|---|---|---|
| Monolithic LiNbO₃ metasurface[36] | 1 GW/cm² | $5\times10^{-5}$ | — |
| LiNbO3 nanodisk on Al substrate[27] | 5.31 GW/cm² | $1.153\times10^{-5}$ | $7.68\times10^{-7}$ |
| LiNbO₃ nanodisk on hyperbolic metamaterial[26] | 5.31 GW/cm² | $5.14\times10^{-5}$ | $1.23\times10^{-6}$ |
| ***LNGW structure (this work)*** | **0.22 mW (1.33GW/cm²)** | **$8.13\times10^{-5}$** | **$1.08\times10^{-5}$** |

## 4 Conclusion

In conclusion, we introduced the breaking symmetry of LiNbO₃ nanograting and waveguide optical system to realize ultrahigh Q factor near BIC, leading to huge local field enhancement. In this way, SH enhancement of ~$10^4$ compared with undersigned LiNbO₃ film and conversion efficiency up to $1.5\times10^{-5}$ with pumping intensity of 1.33GW/cm², is obtained when the optical system has a dissymmetric factor |δ| of 0.2. In particular, we further calculate the SH conversion efficiency exceeding $8.13\times10^{-5}$ at

$|\delta|=0.1$ will be demonstrated. Therefore, we predict that that a giant SH response will be realized when the dissymmetric factor $\delta$ drift approach zero, i.e., from quasi-BICs to BICs. Although our work is developed from theoretical and numerical simulations, our results provide one way for promoting the widespread applications of thin film lithium niobate in integrated nonlinear optics.

*The authors declare no competing interest conflict. H. Lu thanks the support form Distinguishing Talent program of Jinan University and China Scholarship Council (CSC).*

*This work was supported by the National Natural Science Foundation of China (61775084, 61505069, 61705089, and 61705087), Guangdong Special Support Program (2016TQ03X962), and Guangdong Natural Science Funds for Distinguish Young Scholar (2015A030306046), Science Foundation of Guangdong Province (2020A151501791), and Fundamental Research Funds for the Central*